\documentclass[12pt,fleqn]{article}
\usepackage[DIV12]{typearea}
\usepackage{amsmath,amsfonts,amssymb}
\usepackage{graphicx}
\usepackage{cite}
\usepackage{mathrsfs}
\usepackage{pifont}
\usepackage{units}
\usepackage[small,loose]{subfigure}  
\usepackage{color}
\usepackage{cancel}
\usepackage{datetime}

\newcommand{\Eqref}[1]{equation~\eqref{#1}}
\newcommand{\Secref}[1]{section~\ref{#1}}
\newcommand{\Figref}[1]{figure~\ref{#1}}

\newcommand{\eVdist}{\kern-0.06em}

\newcommand{\gev}{\:\text{Ge\eVdist V}}
\newcommand{\tev}{\:\text{Te\eVdist V}}

\newcommand{\D}{\mathrm{d}}

\newcommand{\Hu}{\ensuremath{h_u}}
\newcommand{\Hd}{\ensuremath{h_d}}

\allowdisplaybreaks[1]

\def\mytitle{Hidden SUSY from precision gauge unification}

\title{\mytitle}

\begin{document}

\begin{titlepage}

\begin{flushright}
DESY 13--099\\
TUM--HEP 892/13\\
FLAVOR-EU-43
\end{flushright}

\vspace*{1.0cm}

\begin{center}
{\Large\bf
\mytitle
}

\vspace{1cm}

\textbf{
Sven Krippendorf\footnote[1]{Email: \texttt{krippendorf@th.physik.uni-bonn.de}}{}$^a$,
Hans Peter Nilles\footnote[2]{Email: \texttt{nilles@th.physik.uni-bonn.de}}{}$^a$,
Michael Ratz\footnote[3]{Email: \texttt{michael.ratz@tum.de}}{}$^b$,\\
Martin Wolfgang Winkler\footnote[4]{Email: \texttt{martin.winkler@desy.de}}{}$^c$
}
\\[5mm]
\textit{\small
{}$^a$ Bethe Center for Theoretical Physics\\
{\footnotesize and}\\
Physikalisches Institut der Universit\"at Bonn,\\
Nussallee 12, 53115 Bonn, Germany
}
\\[3mm]
\textit{\small
{}$^b$ Physik--Department T30, Technische Universit\"at M\"unchen, \\
~~James--Franck--Stra\ss e, 85748 Garching, Germany
}
\\[3mm]
\textit{\small
{}$^c$ Deutsches Elektronen--Synchroton DESY, Notkestrasse 85, 22607 Hamburg, Germany.
}
\end{center}

\vspace{1cm}

\begin{abstract}

\noindent
We revisit the implications of naturalness and gauge
unification in the MSSM. We find that precision unification of the couplings in
connection with a small $\mu$ parameter requires a highly compressed
gaugino pattern as it is realized in mirage mediation. Due to the small mass
difference between gluino and LSP, collider limits on the gluino mass are
drastically relaxed. Without further assumptions, the relic density of the LSP
is very close to the observed dark matter density due to coannihilation
effects.

\end{abstract}

\end{titlepage}

\clearpage

\tableofcontents

\section{Introduction}

Two key motivations for studying low--energy supersymmetry are the possibility
for a natural solution to the electroweak hierarchy
problem~\cite{Witten:1981nf} and gauge coupling
unification~\cite{Dimopoulos:1981yj}. Given this input of a natural solution to
the electroweak hierarchy problem and gauge coupling unification, we analyze in
this article their effect on the spectrum of supersymmetric soft masses. Here we
concentrate on the minimal supersymmetric extension of the standard
model (MSSM).

If one ponders about the question whether or not gauge couplings really unify in
the MSSM, it turns out that there is a non--negligible dependence on the pattern
of supersymmetry breaking. In particular, in some of the most popular scenarios,
such as the CMSSM, gauge coupling unification is not precise given
soft masses around the weak scale. This refers to the fact that the strong
coupling  $\alpha_3=g_3^2/(4\pi)$ turns out to be about $3\,\%$ smaller than
$\alpha_1$ and $\alpha_2$ at  $M_\mathrm{GUT}$, which is defined as the scale
where
$\alpha_1$ and $\alpha_2$ unify.

Of course, this discrepancy of the couplings at $M_\mathrm{GUT}$ may originate
from some thresholds at the high scale \cite{Alciati:2005ur} (see also
\cite[S.~Raby,~`Grand unified theories']{Amsler:2008zzb}). However, here we would
like to discuss precision gauge coupling unification (PGU) in the absence of
high--scale threshold corrections\footnote{The size of these threshold
corrections is ultimately set by the UV theory of choice. For a bottom--up analysis we take
it as a UV parameter which we here assume to be vanishing.} and determine the
consequences for the spectrum of soft masses. 

Several examples of spectra achieving PGU have been discussed in the
literature~\cite{Carena:1993qz,Roszkowski:1995cn,Raby:2009sf}. Most of these are
now excluded by the non--observation of superpartners at the LHC or by the
relatively large mass of the Higgs boson. The consideration of naturalness leads
us to the choice of a small $\mu$ parameter which (further) narrows down viable
soft mass spectra. Remarkably, we are led to a highly compressed pattern of
gaugino and higgsino masses at or below the TeV scale, while the scalar
superpartners can be out of reach for the LHC experiment. As the LHC sensitivity
is drastically reduced in case of a small mass splitting between gluino and the
lightest supersymmetric particle (LSP), PGU provides an attractive explanation
on why SUSY has not been discovered so far.

The required spectrum of gaugino masses can be accommodated in
mirage mediation and as such is directly connected to UV scenarios arising
within string
theory~\cite{Choi:2005ge,Choi:2005uz,Falkowski:2005ck,Lebedev:2006qq,Lowen:2008fm}.
Fixing the ratio of gaugino masses with PGU opens interesting windows towards
explicit UV realizations of mirage mediation. 

Further, the problem of dark matter overproduction --- usually
arising in SUSY models with a bino LSP --- can naturally be solved by a
compressed gaugino spectrum. We find that in mirage mediation, PGU is intimately
linked to the occurrence of coannihilations. These allow for a consistent
explanation of dark matter in terms of the lightest neutralino.

This article is organized as follows: in \Secref{sec:pguinmssm} we perform
a model--independent analytical discussion of gauge coupling unification. In
\Secref{sec:3}, we study whether or not PGU can be achieved in realistic
SUSY models taking the CMSSM and mirage mediation as examples. Then, we turn to
the LHC and dark matter phenomenology of models with successful PGU in
\Secref{sec:4}, before concluding in \Secref{sec:conclusions}.

\section{The effective SUSY threshold scale}
\label{sec:pguinmssm}

The MSSM gauge couplings evaluated at the scale of grand unification can be
written as~\cite{Carena:1993ag}
\begin{equation}
 \frac{1}{g_i^2(M_\text{GUT})} ~=~
 \frac{1}{g_i^2(M_Z)} - \frac{b_i^\text{MSSM}}{8\pi^2}\ln\left( \frac{M_\text{GUT}}{M_Z}\right) +  \frac{1}{g_{i,\text{Thr}}^2}  
 +\Delta_i
\end{equation}
where the $b_i^{\rm MSSM}$ denote the standard MSSM beta function
coefficients, $(b_{1},b_{2},b_{3})=(33/5,1,-3)$, while
$\Delta_i$ includes two--loop effects, threshold corrections related to the
heavy standard model fields as well as a possible term required for the
transition between renormalization schemes --- the gauge couplings $g_i^2(M_Z)$
are typically given in the $\overline{\text{MS}}$ rather than the
$\overline{\text{DR}}$ scheme. The threshold corrections within the MSSM
read~\cite{Carena:1993ag}
\begin{equation}\label{eq:MSSMthreshold}
\frac{1}{g_{i,\text{Thr}}^2}~=~\sum\limits_{\eta}  \frac{b_i^\eta}{8\pi^2}\,\ln\left( \frac{M_\eta}{M_Z} \right)\;,
\end{equation}
where the sum runs over all sparticles and heavy Higgs fields with mass $M_\eta$
evaluated at the low scale and $b_i^\eta$ denoting the contribution of the
particle $\eta$ to the $i^\mathrm{th}$ $\beta$--function coefficient
(cf.~\cite{Lahanas:1994dj} for the exact expressions of $b_i^\eta$ in the
MSSM).\footnote{Here we neglect two--loop contributions to the
thresholds.}

To arrive at a consistent picture of unification, the $g_i$ are required
to meet in a single point. Assuming the absence of threshold corrections induced
by heavy fields of a grand unified theory (GUT), this implies that the
masses of the MSSM fields must arrange such that their overall threshold to the
$g_i$ allows for PGU. 

To get a first impression on the effect of soft masses on gauge coupling
unification, let us make the ad hoc assumption that all sparticles as well as the heavy Higgs doublet have a common
mass $T_\text{SUSY}$ which we call the SUSY threshold scale. In this case, the overall
threshold correction~\eqref{eq:MSSMthreshold} to the gauge couplings takes a
very simple form. We find
\begin{equation}
 \frac{1}{g_{i,\text{Thr}}^2}~=~ 
 \frac{b_i^\text{MSSM}-b_i^\text{SM}}{8\pi^2}\,\ln\left( \frac{T_\text{SUSY}}{M_Z} \right)\;,
\end{equation}
where $b_i^\text{SM}$ stands for the beta function coefficients of the standard
model. Now we can determine the SUSY threshold scale for which the gauge couplings meet
exactly. To quantify the deviation from PGU, we
follow~\cite{Raby:2009sf,Anandakrishnan:2013cwa} and introduce
\begin{equation}
 \epsilon_3~=~\frac{g_3^2(M_\text{GUT})-g_{1,2}^2(M_\text{GUT})}{g_{1,2}^2(M_\text{GUT})}\;,
\end{equation}
where we define $M_\text{GUT}$ as the scale at which $g_1$ and $g_2$ meet.

\begin{figure}[htp]
\begin{center}
\includegraphics[width=9cm]{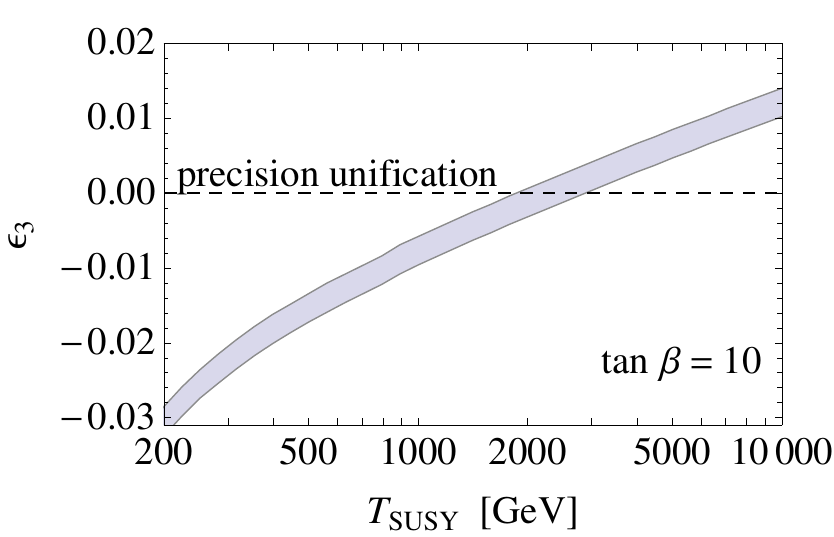}
\end{center}
\caption{\footnotesize{Deviation from PGU as a
function of $T_\text{SUSY}$. The width of the band corresponds
to the $1\,\sigma$ experimental error in $\alpha_s(M_Z)$.}}
\label{fig:SUSYscale}
\end{figure}

In \Figref{fig:SUSYscale} we depict $\epsilon_3$ as a
function of the SUSY threshold scale. For the determination of $\epsilon_3$, we
have used Softsusy (version 3.3.2)~\cite{Allanach:2001kg}, where we set the
strong coupling strength to $\alpha_s(M_Z)= 0.1184\pm
0.0007$~\cite{Bethke:2012jm} and $\tan\beta=10$. It can be seen that the three
gauge couplings would exactly meet in a single point if all superpartners as
well as the heavy Higgs doublet had a common mass $T_\text{SUSY}\simeq 2\tev$.
Note that the required value of $T_\text{SUSY}$ has a very mild dependence on
$\tan\beta$ which only affects the gauge coupling at the two--loop level. 

Turning to realistic SUSY models , despite the fact that not all superpartners
have a common mass, it is still possible to define an effective SUSY threshold
scale $T_\text{SUSY}$ through the relation
\begin{equation}
\frac{b_i^\text{MSSM}-b_i^\text{SM}}{8\pi^2}\,\ln\left( \frac{T_\text{SUSY}}{M_Z}\right)
~\equiv~
\sum\limits_{\eta}  \frac{b_i^\eta}{8\pi^2}\,\ln\left( \frac{M_\eta}{M_Z}\right)  
+ \left(c_1 + c_2\,b_i^\text{MSSM}\right)
\end{equation}
with the same $(T_\text{SUSY},c_1,c_2)$ for all three
standard model gauge groups. Note that the term $c_1$ accounts for an
overall shift of all $g_i(M_\text{GUT})$, while the $c_2$ term
results in a change of $M_\text{GUT}$. 

Therefore $T_\text{SUSY}$ has a very simple interpretation: a SUSY spectrum with
an effective SUSY threshold scale $T_\text{SUSY}$ has the same effect on gauge
coupling unification as a degenerate spectrum with all superpartners having this
common mass, up to a change of $M_\text{GUT}$ and/or the unified gauge coupling.

The effective SUSY threshold scale takes the form\footnote{A simplified version of this formula was discussed in~\cite{Carena:1993ag}.}
\begin{equation}\label{eq:effSUSYscale}
 T_\text{SUSY}~=~
 \frac{m_{\widetilde{W}}^{32/19}\,m_{\widetilde{h}}^{12/19}\,m_H^{3/19}}{m_{\widetilde{g}}^{28/19}}\;X_\text{sfermion}\; ,
\end{equation}
where 
$m_{\widetilde{W}}$, $m_{\widetilde{h}}$, $m_H$ and $m_{\widetilde{g}}$ denote
the the mass of the wino, the light and heavy MSSM Higgs and gluino,
respectively. The sfermion contribution reads
\begin{equation}\label{eq:sfermionthreshold}
 X_\text{sfermion}
 ~=~\prod\limits_{i=1\dots3}
 \left(\frac{m_{\widetilde{L}^{(i)}}^{3/19}}{m_{\widetilde{D}^{(i)}}^{3/19}}\right)
 \left(\frac{{m_{\widetilde{Q}_\mathrm{L}^{(i)}}^{7/19}}}{
 m_{\widetilde{E}^{(i)}}^{2/19}\,
 m_{\widetilde{U}^{(i)
 }}^{5/19}}\right)
 \;,
\end{equation}
where the masses of the sfermions appear in a self--explanatory notation. As we will show explicitly in the next section, $X_\text{sfermion}\simeq 1$ whenever
the sfermion masses are universal within a SU(5) multiplet at the high scale which is expected to
hold in a GUT model. The mass splitting of sleptons and
squarks through renormalization group running does not affect this conclusion. Hence, the sfermions leave no direct effect
on PGU.

Note that the insensitivity of gauge coupling unification to the
scale of sfermion masses is due to the completeness of these GUT multiplets.
Only split GUT multiplets give non--trivial contributions to $T_\text{SUSY}$
(cf.~\Eqref{eq:effSUSYscale}). This is also utilized in models
of split supersymmetry~\cite{ArkaniHamed:2004fb}. It is also interesting that
the decoupling of complete generations of soft scalar masses, e.g.\ the first and second generation, 
does not affect PGU, allowing for a realization of ``natural
supersymmetry'' with a light third generation of soft scalar masses.

\section{Gauge coupling unification in MSSM models}
\label{sec:3}

As can be seen from~\Eqref{eq:effSUSYscale}, $T_\text{SUSY}$ mainly
depends on the higgsino mass and on the mass ratio
$m_{\widetilde{W}}\,:\,m_{\widetilde{g}}$.  In the following, after discussing
the case of uncompressed gaugino masses, we shall show how a change in this mass
ratio can lead to PGU.

\subsection{Uncompressed gaugino mass spectra}

\begin{figure}[htp]
\begin{center}
\includegraphics[width=9cm]{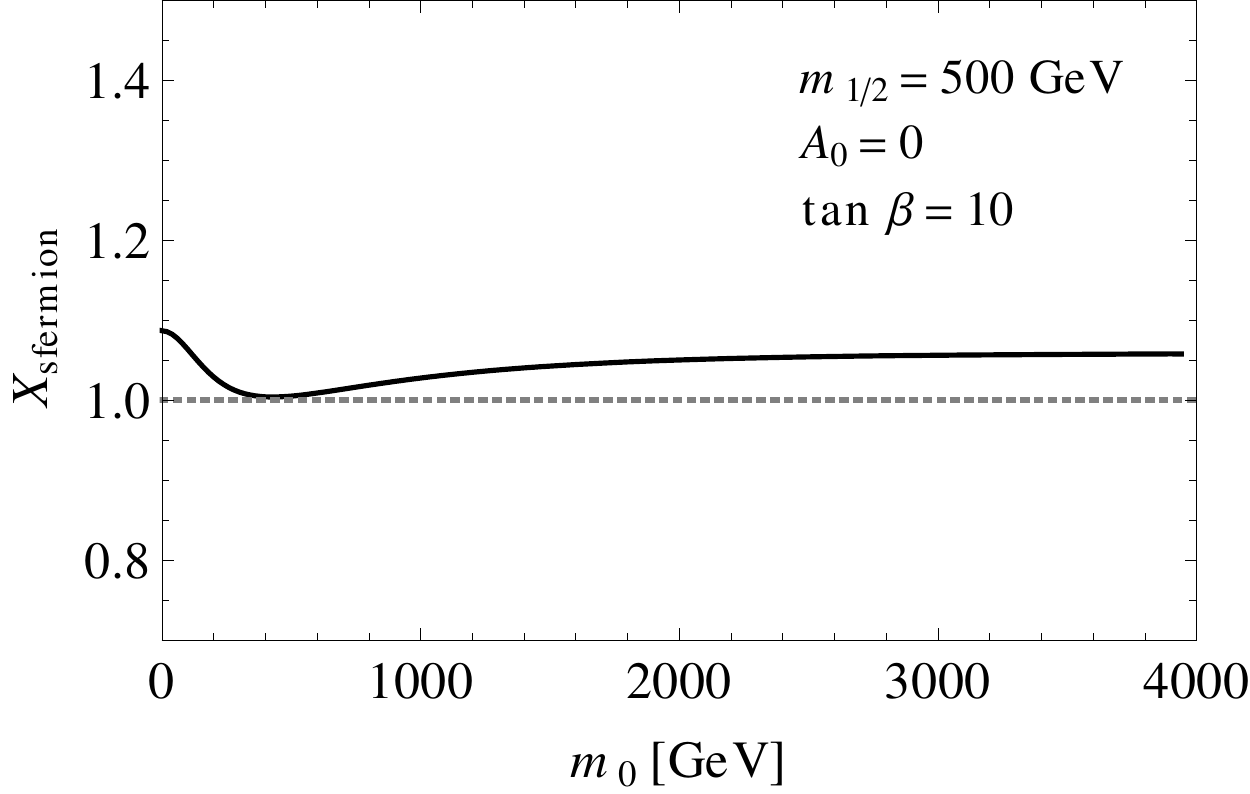}
\end{center}
\caption{\footnotesize{Effect of sfermions on gauge coupling unification in the CMSSM for varying $m_0$.}}
\label{fig:xsfermion}
\end{figure}

In models with gravity mediated SUSY breaking, the gaugino masses
are typically expected to unify at the GUT scale $M_\text{GUT}$. 
Taking $m_{1/2}$ to be the universal gaugino mass at the GUT scale, the physical
mass of wino and gluino can be approximated as
\begin{equation}
 m_{\widetilde{W}} ~\simeq~ 0.9\,m_{1/2}
 \quad\text{and}\quad m_{\widetilde{g}} ~\simeq~ 2.5\,m_{1/2}\;.
\end{equation}
In turn the effective SUSY threshold scale reads
\begin{equation}
 T_\text{SUSY} 
 ~\simeq~ 
 0.3\, \left( m_{\widetilde{h}}^{12} \, m_{1/2}^{4}\, 
 m_{H}^{3} \right)^{1/19}\,X_\text{sfermion}\;.
\end{equation}
In the limit where the sfermions are substantially heavier than the gauginos,
sfermion masses are only weakly affected by RGE running. As long as the sfermion
masses are universal within SU(5) multiplets at the high scale, we find
$X_\text{sfermion}\simeq 1$. Even if gaugino masses are non--negligible and the
masses of squarks and sleptons are split by RGE running, their effect on gauge
coupling unification remains small. This is illustrated in
\Figref{fig:xsfermion} where we depict $X_\text{sfermion}$ in the CMSSM for
a fixed gaugino mass $m_{1/2}$ and a varying scalar mass $m_0$. It can be seen
that $X_\text{sfermion}$ never deviates by more than a few per cent from one.

Taking $X_\text{sfermion}= 1$, we can determine the higgsino mass which is
required to obtain PGU, i.e. $T_\text{SUSY}\simeq 2\tev$. We find
\begin{equation}
 m_{\widetilde{h}}~\simeq~
 20\tev \times \left(\frac{\text{TeV}}{m_{1/2}}\right)^{1/3} \, 
 \left(\frac{\text{TeV}}{m_H}\right)^{1/4}\; .
\end{equation}
Due to the small exponents of the last two terms, higgsino masses as large as
$m_{\widetilde{h}}=\mathcal{O}(10\tev)$ are required for gauge coupling
unification, even if we take $m_{1/2}$ and $m_H$ in the multi--TeV range. As is
well known, such heavy higgsinos lead to `unnaturally' large fine--tuning as can
be seen from the mass of the $Z$ boson which --- at tree level --- can be written as
$M_Z \simeq - 2\,m_{\Hu}^2 - 2\,|\mu|^2$ for $\tan\beta\gg 1$. In the case of
a large higgsino mass, severe cancellations between $m_{\Hu}$ and $\mu$ are
required.

We will therefore now turn to a class of models in which PGU can be accommodated
more naturally.

\subsection{Compressed gaugino mass spectra}\label{sec:Compressed}

The effective $T_\mathrm{SUSY}$ can be enhanced if one increases the wino mass relative
to the gluino mass. Specifically, we now discuss the possibility of realizing
PGU by non--universal gaugino masses which can be accommodated in soft mass
spectra arising from UV completions of the MSSM. To obtain non--universal gaugino
masses, an obvious possibility is to use anomaly mediated contributions. While
in pure anomaly mediation the hierarchy between the gluino and wino is even
larger than in gravity mediation, more compressed gaugino spectra can be
obtained in schemes with mixed gravity/anomaly mediation, also referred to as
mirage scheme.

At first sight, having gravity and anomaly mediation at similar strength appears
as an ad hoc assumption. But note that this naturally occurs in a wide class of
string models due to the interplay between moduli stabilization and SUSY
breaking. Examples of such mixed mediation schemes have been discussed in the
context of type IIB string theory with moduli stabilization following
KKLT~\cite{Kachru:2003aw,Choi:2004sx,Choi:2005ge,Choi:2005uz,Lebedev:2006qq,Dudas:2012wi},
in the context of heterotic string theory~\cite{Lowen:2008fm,Krippendorf:2012ir,Badziak:2012yg,Kaufman:2013pya}, and in the context of
$G_2$ compactifications~\cite{Acharya:2012tw}.

In mirage mediation, the gaugino masses at the high scale can be written as
\begin{equation}
 M_i~=~ \frac{m_{3/2}}{16\,\pi^2}\,\left(\varrho+b_i^\text{MSSM}\,g^2\right)\;,
\end{equation}
where $m_{3/2}$ denotes the gravitino mass and $\rho$ parametrizes the
gravity mediated contribution to the gaugino masses, in the limit
of vanishing $\varrho$ pure anomaly mediation is recovered. Notice that $|M_3| <
|M_2|$ at the GUT scale unless for $\varrho \lesssim 0.5$. The gaugino masses at
the low scale can be approximated as
\begin{subequations}\label{eq:gauginomiragelow}
\begin{eqnarray}
 m_{\widetilde{B}} &\simeq& 0.45\,(\varrho+3.3)\;\frac{m_{3/2}}{16\pi^2}\;,\\
 m_{\widetilde{W}} &\simeq& 0.9\,(\varrho+0.5)\;\frac{m_{3/2}}{16\pi^2}\;,\\
 m_{\widetilde{g}} &\simeq& 2.4\,(\varrho-1.5)\;\frac{m_{3/2}}{16\pi^2}\;.
\end{eqnarray}
\end{subequations}
The gaugino spectrum is thus typically more compressed than in models with
gaugino mass unification. In the regime $0.8\lesssim\varrho\lesssim 2.5$
the gluino is even the lightest gaugino.

In the context of mirage mediation, the scalar masses and trilinear terms are
model dependent. The former are of the order of the gravitino mass unless in
case of sequestering. The trilinear soft terms are typically suppressed
compared to the gravitino mass and can thus be neglected. In this study we
assume sfermion masses in the multi--TeV regime as suggested by the KKLT and
heterotic models with $F$--term uplifting~\cite{Lebedev:2006qq,Abe:2006xp,Kallosh:2006dv,Lowen:2008fm}. 

In order to avoid a naturalness problem, we should require that
$m_{h_{u,d}} = 
m_{\widetilde{Q}_{L}^{(3)}} 
= m_{{\widetilde{U}^{(3)}}}$ approximately
holds a the GUT scale.\footnote{This is a very reasonable assumption as e.g.\ in
the heterotic models, these fields have the same localization properties (see
discussion in~\cite{Krippendorf:2012ir,Badziak:2012yg}).} In this case, the RGE
trajectory of $m_{\Hu}$ exhibits the well known focus
point~\cite{Feng:1999mn,Feng:1999zg} such that multi--TeV scalars are not
unnatural. In the focus point scenario $m_{\Hu}$ is driven to a very small value
at the low scale, while $m_{\Hd}$ is not considerably affected by RGE running.
This implies $m_H \simeq m_{\Hd}$ for the physical mass of the heavy MSSM Higgs
as well as a suppressed $\mu$ term~\cite{Feng:1999mn}. Therefore, the
above assumption is equivalent to $ m_H = m_{\widetilde{Q}_{L}^{(3)}} =
m_{\widetilde{U}^{(3)}} \gg\mu$ up to sub--leading corrections. As the running of the
gauge couplings depends on $\mu$ rather than the Higgs soft masses this form of
the input is more suitable for our purposes. To be specific we set
\begin{equation}
 m_H~=~m_i ~\equiv~ m_0\;,
\end{equation}
where $m_i$ denotes the mass of the sfermion $i$ at the high scale. We are thus
left with the five free parameters $m_{3/2}$, $\varrho$, $m_0$, $\mu$ and
$\tan\beta$. To avoid excessive fine--tuning we restrict our attention to
$\mu\leq 2\tev$. Further, for any point in the parameter space, we eliminate
$m_0$ by requiring that the mass of the light MSSM Higgs is $m_h=126\gev$, which
typically leads to $m_0 \sim 15\tev$. We should point out that our conclusions
are insensitive to the assumptions in the scalar sector as the latter hardly
affect gauge coupling unification. Thus our discussions is also valid in schemes
where some of the complete sfermion families receive a mass $\sim m_{3/2}$
which typically is larger than $m_0$.

\begin{figure}[h]
\begin{center}
\includegraphics[width=9cm]{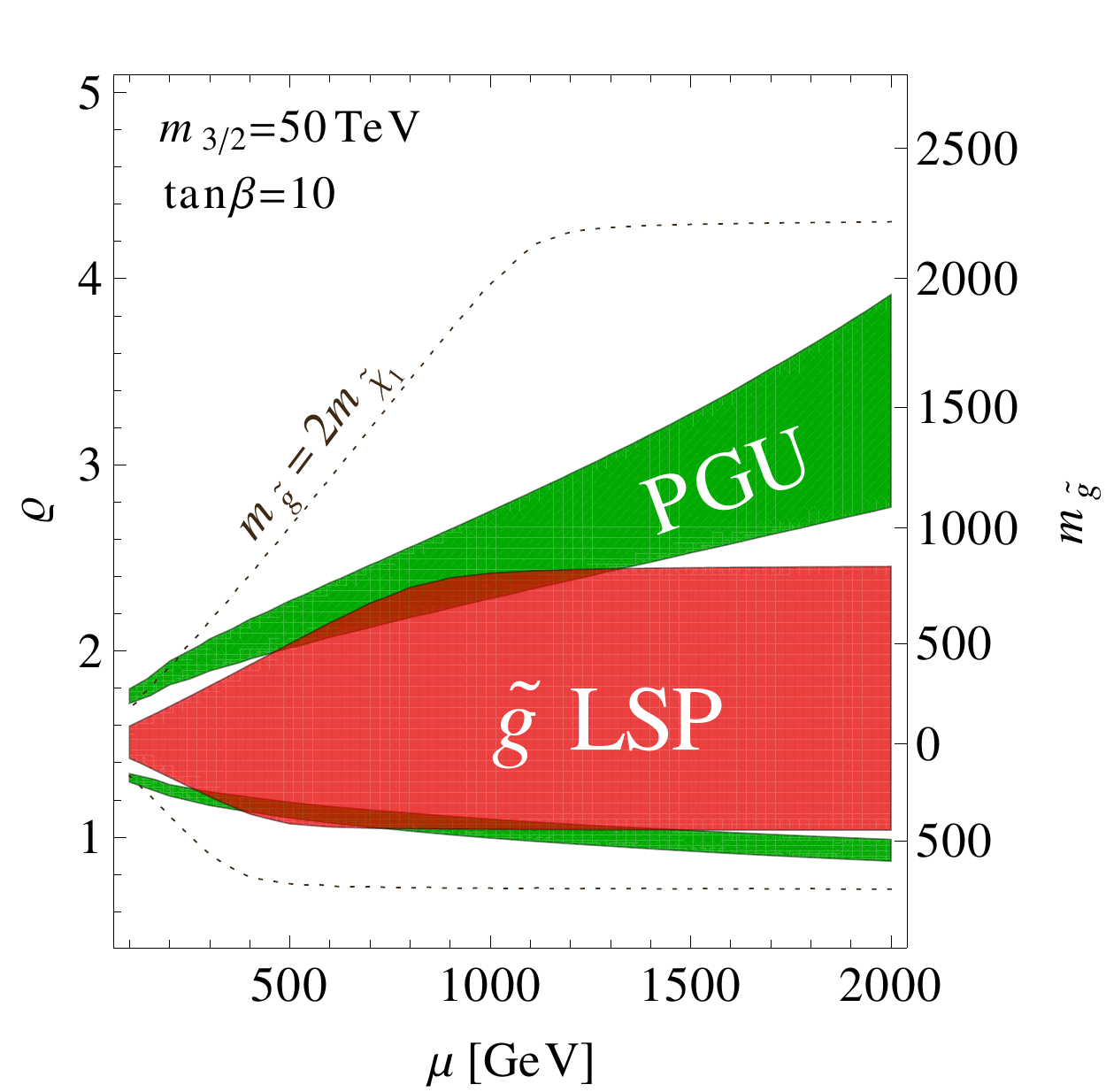}
\end{center}
\caption{\footnotesize{Parameter scan in the $(\mu,\varrho)$-plane for fixed
$m_{3/2}=50\tev$. The region where the gauge couplings unify precisely within
the experimental error on the strong coupling are shown in green. The red region
exhibits a gluino LSP. The dotted contour indicates where the mass ratio between
gluino and LSP becomes two.}}
\label{fig:Mirage}
\end{figure}

We now can ask what constraints on the parameter space arise from imposing PGU.
In \Figref{fig:Mirage}, we present the parameter region consistent with PGU
for fixed $m_{3/2}$ and $\tan\beta$. Precision unification clearly favors a
compressed spectrum, i.e.\ a spectrum where the mass difference between gluino
and LSP is small.
For $1.1 < \varrho < 2.4$, the gluino is the LSP, unless the higgsino becomes
even lighter (shrinking of the red band at the left side in
\Figref{fig:Mirage}). The two regions in parameter space consistent with PGU
correspond to the two regions where the mass ratio of wino and gluino gives
rise to the required effective SUSY threshold scale.  

There is only a weak dependence on the overall scale of gaugino masses which is
set by $m_{3/2}$. Therefore, our results remain qualitatively unchanged for
different choices of $m_{3/2}$. Similarly, PGU only shows a weak dependence on
$\tan\beta$. For illustration, we have included scans with varying $m_{3/2}$ and $\tan\beta$ in appendix~\ref{sec:moreplots}.

\section{Implications for the LHC and dark matter}
\label{sec:4}

\subsection{LHC discovery potential}

We can now turn to the implications of PGU for SUSY searches at the LHC.  As
described in the previous section, this naturally leads us to models with mirage
mediation where PGU can naturally be accommodated with a small $\mu$.

Within the mirage scheme, solid predictions on the superpartner spectrum can be
made. While the scalar superpartners are typically outside the reach of the LHC,
the gauginos and higgsinos remain relatively light. We cannot predict the
hierarchy among gauginos and higgsinos, but there is a general trend that PGU
prefers a spectrum in which the mass difference between the gluino and the
lightest neutralino is small.

 Such compressed SUSY models are much more difficult to detect at the LHC than
ordinary SUSY models like the CMSSM (see
e.g.~\cite{LeCompte:2011cn,LeCompte:2011fh,Dreiner:2012gx}). While there is
still a strong gluino pair production, the subsequent decay of gluinos yields
only a small amount of visible energy. In the experimental searches, a sizable
fraction of the potential signal events is rejected due to too soft jets. In the
case of extreme compression, only events with initial state radiation can pass
the event cuts, and the signal acceptance is typically well below
$1\%$~\cite{LeCompte:2011fh}.

If the gluino is the next--to--lightest supersymmetric particle, the relevant gluino decay modes read
\begin{subequations}
\begin{eqnarray}
  \widetilde{g} &\rightarrow&  q\bar{q} + \widetilde{\chi}_1\;,\label{eq:gluinoquark}\\
  \widetilde{g} &\rightarrow&  g + \widetilde{\chi}_1\;.\label{eq:gluinogluon}
\end{eqnarray}
\end{subequations}
where $\widetilde{\chi}_1$ denotes the lightest neutralino. The first process
proceeds via an off--shell squark, the second via a quark / squark loop.
Additional decay modes open up if further neutralinos or charginos are lighter
than the gluino.

It is instructive to determine the regime of gluino and (lightest) neutralino
masses preferred by PGU. For this purpose, we have generated a large data sample
with random choice of input parameters in the intervals
\begin{equation}\label{eq:boundaries}
 \varrho=0.5-30\;,\quad m_{3/2}=\frac{40-200\tev}{\varrho}\;,  \quad \mu=0.1-2\tev\;, \quad \tan\beta=10-50\;.
\end{equation}
Note that the considered range of $m_{3/2}$ corresponds to a gravity
mediated contribution to the gaugino masses of $0.25-1.25\tev$.

\begin{figure}[h!]
\begin{center}
\includegraphics[width=11cm]{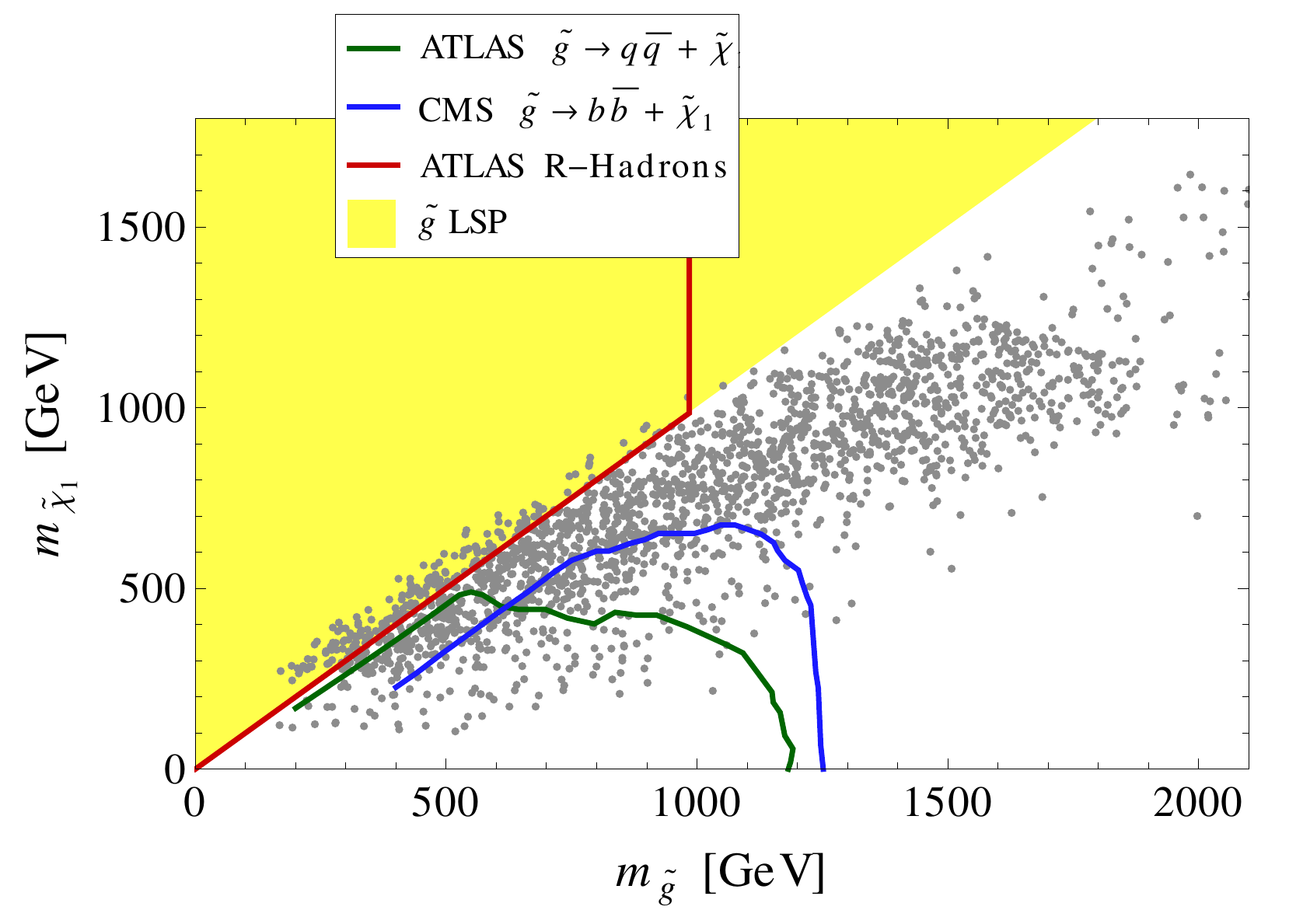}
\end{center}
\caption{\footnotesize{Parameter points with successful gauge coupling
unification (gray). Also shown are the constraints on the gluino mass for the
decay modes $\widetilde{g} \rightarrow  q\bar{q} + \widetilde{\chi}_1$ and
$\widetilde{g} \rightarrow  b\bar{b} + \widetilde{\chi}_1$ by ATLAS and CMS (see
text). The yellow region features a gluino LSP which is constrained by the ATLAS
search for stable R--hadrons.}}
\label{fig:scatter}
\end{figure}

In \Figref{fig:scatter}, we provide a scatter plot of those parameter sets
which lead to successful PGU in the
($m_{\widetilde{g}},m_{\widetilde{\chi}_1}$)--plane. To guide the eye, we have
also included the present limits on the gluino mass from the ATLAS search for
jets plus missing energy at $\int \! \D t\, L =5.8\,
\text{fb}^{-1}$~\cite{ATLAS:2012ona} and from the CMS search for $b$--jets plus
missing energy at $\int \! \D t\, L =19.4\,
\text{fb}^{-1}$~\cite{CMS:wwa}.\footnote{See~\cite{ATLAS:2012ksq} for
a similar analysis by ATLAS.} Both analyses are based on a simplified model with
just the gluino and the LSP assuming $100\%$ branching of the
process~\eqref{eq:gluinoquark}. CMS additionally requires the final state quarks
to be bottoms. Both limits are not strictly applicable as the considered decay
modes do not necessarily dominate in the set--up we consider. In particular, we
may encounter longer decay chains if the gluino decay to heavier neutralinos
/ charginos is kinematically accessible, leading to events with more and softer
jets than in the simplified model. As too soft jets typically fail the cuts
performed in the ATLAS and CMS analyses, this can clearly affect the
constraints. However, very soft jets would mainly arise in parameter regions
with a strongly compressed spectrum where initial state radiation is anyway
required to pass the event trigger, the latter being insensitive to the gluino
decay. Therefore, we believe that at least the ATLAS limit can still be used as
a reasonable estimate. The stronger CMS limit should not be applied, as we do
not find a preference for $b$--jets in the final state compared to light quark or
gluon jets. We have, nevertheless, included it in \Figref{fig:scatter} as
the LHC gluon searches without $b$--tagging should reach a similar sensitivity once
the full data sample collected at $\sqrt{s}=8\tev$ is analyzed. Therefore, the
CMS constraint can be seen as a projection of the current sensitivity of the LHC
to gluinos.

A small fraction of the benchmark points features a gluino LSP. Apart from the
fact that a gluino LSP is very unfavorable from a cosmological perspective,
stable gluinos with $m_{\widetilde{g}}< 985\gev$ are excluded by the ATLAS
search for R--hadrons~\cite{Aad:2012pra}. This constraint is also depicted in
\Figref{fig:scatter}. 

We observe that more than $90\%$ of the benchmark points with PGU fulfill $m_{\widetilde{\chi}_1}>0.5\,m_{\widetilde{g}}$. The strong
compression of the spectrum implies that the limits on the gluino mass are
considerably weaker than in ordinary SUSY schemes like the CMSSM. In particular
we find that a large fraction of the benchmark points with
$m_{\widetilde{g}}=500-1000\gev$ is not excluded by the LHC.

We should also point out that the gluino can be rather long--lived in the
scenario we consider. This is partly because of the phase space suppression due
to the small mass difference between gluino and neutralino, and partly because
of the large mass of the squarks which mediate the gluino decay. The gluino
decay rates in case of $m_{\widetilde{\chi}_1} \sim m_{\widetilde{g}}$ scale as
(see e.g.~\cite{Barbieri:1987ed})
\begin{subequations}\label{eq:decayrates}
\begin{eqnarray}
 \Gamma(\widetilde{g} \rightarrow  q\bar{q} + \widetilde{\chi}_1) &\propto& \text{gaugino fraction}\times \frac{\left(m_{\widetilde{g}}-m_{ \widetilde{\chi}_1}\right)^5}{m_{\widetilde{q}}^4}\;,\\
\qquad  \Gamma(\widetilde{g} \rightarrow  g + \widetilde{\chi}_1) &\propto& \text{higgsino fraction}\times \frac{\left(m_{\widetilde{g}}-m_{ \widetilde{\chi}_1}\right)^3\,m_t^2}{m_{\widetilde{q}}^4}\;,
\end{eqnarray} 
\end{subequations}
where $m_t$ denotes the top mass and $m_{\widetilde{q}}$ the mass of the squark
in the intermediate state. In the set--up we consider, the squarks are in the range $m_{\widetilde{q}}\sim 15\tev$ as implied by the requirement
$m_h=126\gev$ (see \Secref{sec:Compressed}). Note that the decay of the
gluino into two quarks scales with the gaugino fraction of $\widetilde{\chi}_1$,
as the higgsino coupling to light quarks is suppressed, while top quarks in the
final state are kinematically inaccessible for $m_{\widetilde{\chi}_1} \sim
m_{\widetilde{g}}$. The decay into gluon and neutralino, on the other hand, scales
with the higgsino component of $\widetilde{\chi}_1$. In our set--up the mass
splitting between left- and right--handed squarks is typically small, i.e.\ parity
is approximately preserved in the squark sector. Therefore, the decay of the
gluino into gaugino and gluon --- which requires parity
violation~\cite{Haber:1983fc} --- can be neglected.

\begin{figure}[h!]
\begin{center}
\includegraphics[width=7.5cm]{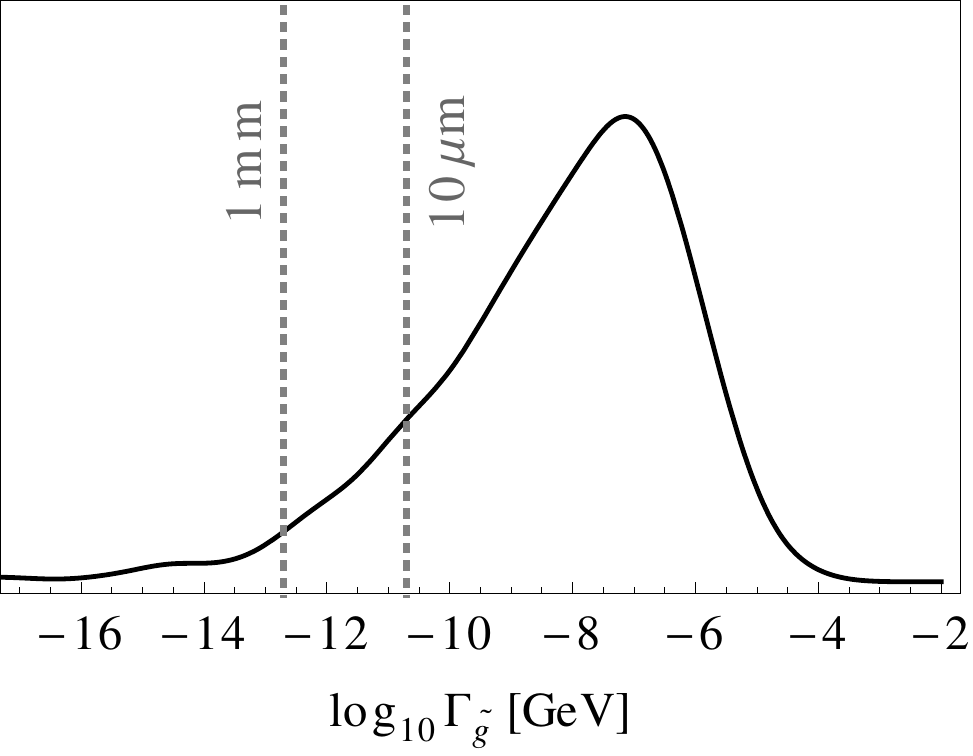}
\end{center}
\caption{\footnotesize{Distribution of the gluino decay rate within the
benchmark sample. The vertical dashed lines indicate where the gluino decay length
$c/\Gamma_{\widetilde{g}}$ reaches $10\:\mu\text{m}$ (roughly the LHC
resolution) and $1\:\text{mm}$ respectively.}}
\label{fig:gluinorate}
\end{figure}

We have systematically determined the total gluino decay rate
$\Gamma_{\widetilde{g}}$ for the benchmark sample using the tool SDECAY
1.3~\cite{Muhlleitner:2003vg}, the corresponding distribution is shown in
\Figref{fig:gluinorate}.\footnote{The distribution was slightly smoothed.}
We find that for slightly more than $10\%$ of the benchmark points the decay
length $c/\Gamma_{\widetilde{g}}$ exceeds $10\,\mu\text{m}$ which roughly corresponds
to the LHC resolution. This suggests that, in a non--negligible fraction of the
parameter space, one might be able to observe displaced vertex signatures. Note
that, in our approach, we assumed universal squark masses at the high scale and
fixed $m_{\widetilde{q}}$ such that the correct Higgs mass is obtained. There
exist, however, theoretical uncertainties in the calculation of the Higgs mass
at the level of a few GeV~\cite{Draper:2013cka}. As the squark mass enters the loop corrections to
$m_h$ only logarithmically this translates into an $\mathcal{O}(1)$ uncertainty
on the squark mass.\footnote{Note also that in the benchmark sample we have set
$\tan\beta=10-50$. For $\tan\beta < 10$, larger squark masses are required in
order to obtain $m_h=126\gev$, i.e. the life--time of the gluino would be further
enhanced.} Given that $\Gamma_{\widetilde{g}}\propto m_{\widetilde{q}}^{-4}$,
our calculation of the gluino decay rate can, thus, only be seen as an order of
magnitude estimate. For many benchmark points $c/\Gamma_{\widetilde{g}}$ is at
least close to the LHC position resolution. Therefore, we believe that there is
indeed a realistic chance to observe displaced vertices. Particularly long
gluino life--times are obtained if the LSP is gaugino--like. In this case
$\Gamma_{\widetilde{g}}$ is suppressed by the fifth power of the mass splitting
(or the decay must proceed through the subleading higgsino admixture of
$\widetilde{\chi}_1$).

The gluino will, however, almost certainly decay within the inner pixel
detectors of ATLAS / CMS. It is unlikely that the gluino decay vertex is more
than a few mm from the initial collision vertex (see
\Figref{fig:gluinorate}). While dedicated searches for gluino decays with
such tiny displacements do currently not exist, it seems feasible to employ
similar methods as for the identification of $b$--jets. In particular one could
study the distribution of the transverse impact parameter in events with jets
and missing energy. One might worry that the jets arising from the gluino decays
are too soft to be detected. However, we find that a decay length above the LHC
resolution can be obtained for mass splittings
$m_{\widetilde{g}}-m_{\widetilde{\chi}_1}$ as large as about $100\gev$
(depending on the composition of $\widetilde{\chi}_1$). We thus want to
emphasize the importance to develop search strategies for gluinos with decays
slightly displaced from the collision vertex. It is also important to determine
how $b$--tagging is affected if the superpartner spectrum contains gluinos with a
decay length similar as that of $b$--mesons.

Even if gluinos are too short--lived to give rise to displaced vertices, the
study of gluino decays is of high interest. This is because details of the
superpartner sector are encrypted in the gluino decay pattern. It will be
difficult to distinguish the multi--jet decay modes of the gluino. However, as
was pointed out in~\cite{Sato:2012xf}, it is feasible to identify the radiative
mode $\widetilde{g} \rightarrow  g + \widetilde{\chi}_1$ which typically gives
rise to events with a low jet multiplicity. This particular decay mode is very
sensitive to the higgsino component of the lightest neutralino
(cf.~\eqref{eq:decayrates}). A large branching fraction in this mode would thus
hint at a relatively small $\mu$ term. Of course, it will be challenging to
determine the full set of high scale parameters just from the gluino decay. But
in any case it is remarkable that gluino decays can in principle be used to
learn about the spectrum of superpartners which are, in principle, outside the
reach of the LHC.

\subsection{Dark Matter}

As it is well known, the lightest neutralino is a very good candidate for the
dark matter in the universe. However, in the simplest realizations of gravity
mediation, the lightest neutralino is either bino--like or higgsino--like.
Considerable neutralino mixing does only occur in some narrow regions of
parameter space. Due to its small annihilation cross section, the bino density
from thermal production typically exceeds the observed dark matter density
$\Omega_{\text{DM}}$ by far. The latter has recently been determined at
high accuracy by the Planck collaboration~\cite{Ade:2013zuv},
\begin{equation}
 \Omega_{\text{DM}}\,h^2~=~0.1196 \pm 0.0031\;.
\end{equation}
Higgsinos, on the other hand, undergo efficient annihilations into third
generation quarks or gauge bosons, coannihilations with the charged higgsinos
further enhance their cross section. Hence, the relic density of a higgsino LSP
is typically well below the dark matter density. In both cases one has to invoke
a non--standard cosmological history if the LSP is to account for the observed
dark matter. In case of the higgsino, non--thermal production e.g.\ by the decay
of a heavy gravitino or modulus field is a viable option (see
e.g.~\cite{Moroi:1999zb,Fujii:2001xp,Kitano:2008tk}). For a bino LSP the
situation is more unfavorable as, if its abundance is diluted by the decay of a
heavy field, binos are typically regenerated by the same decay, reintroducing
precisely the same problem~\cite{Moroi:1994rs,Nakamura:2006uc}.

\begin{figure}[h!]
\begin{center}
\includegraphics[width=7.5cm]{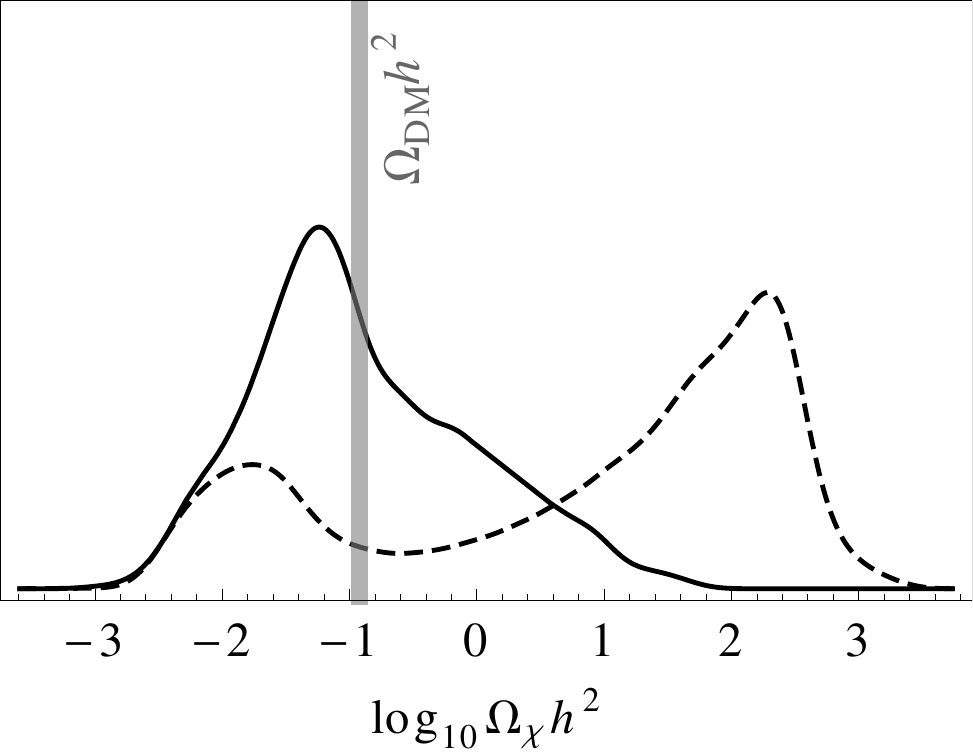}
\end{center}
\caption{\footnotesize{Distribution of the thermal neutralino relic density for
the benchmark sample with (solid) and without (dashed) requiring PGU. The
observed dark matter density is indicated by the gray contour.}}
\label{fig:relicdensity}\end{figure}

In mirage mediation, the gaugino masses are non--universal at the high scale due
to the anomaly--mediated contributions. We have seen that if we require PGU, we
are drawn into a region of parameter space were the gaugino spectrum is highly
compressed. This turns out to be very favorable for the dark matter density. We
have determined the neutralino relic density $\Omega_{\chi}\,h^2$ for a large
benchmark sample using the boundaries~\eqref{eq:boundaries} with the tool
MicrOMEGAs~2.4.5~\cite{Belanger:2006is}. In \Figref{fig:relicdensity} we
compare the distribution of $\Omega_{\chi}\,h^2$ among the sample points with
and without imposing PGU.\footnote{For this figure we have disregarded benchmark
points with a gluino LSP. The distribution was again slightly smoothed.}

We find that without the requirement of PGU, the lightest neutralino is
typically bino-- or higgsino--like, similar as in the models with gaugino mass
unification. The distribution of the relic density among the benchmark points
has two peaks corresponding to the bino and the higgsino case respectively. 

If we include the requirement of PGU, the picture changes dramatically.
Remarkably, as most benchmark points with PGU feature a compressed spectrum, the
distribution is now peaked close to observed dark matter density.  The reason is
that in the parameter regions with precision unification, the mass difference
between bino and wino is typically $\sim 10\%$, suggesting the occurrence of
coannihilations. In particular in case of a bino LSP, wino pair processes would
still dominate the annihilation cross section. These would be suppressed by a
Boltzmann factor 
\begin{equation}
 B~=~ \exp{\left(-2\, \frac{m_{\widetilde{W}}-m_{\widetilde{B}}}{T_F}\right)}\,
\end{equation}
with the freeze--out temperature $T_F \sim  m_{\widetilde{B}}/20$. Given a
splitting of $\sim 10\%$ one finds $B\sim 0.01$. This factor compensates for the
large wino pair-annihilation cross section, yielding a relic dark matter
density consistent with observation. Another possibility to obtain the correct
relic density is by mixing effects in the neutralino sector.

\begin{figure}[h!]
\begin{center}
\includegraphics[width=12cm]{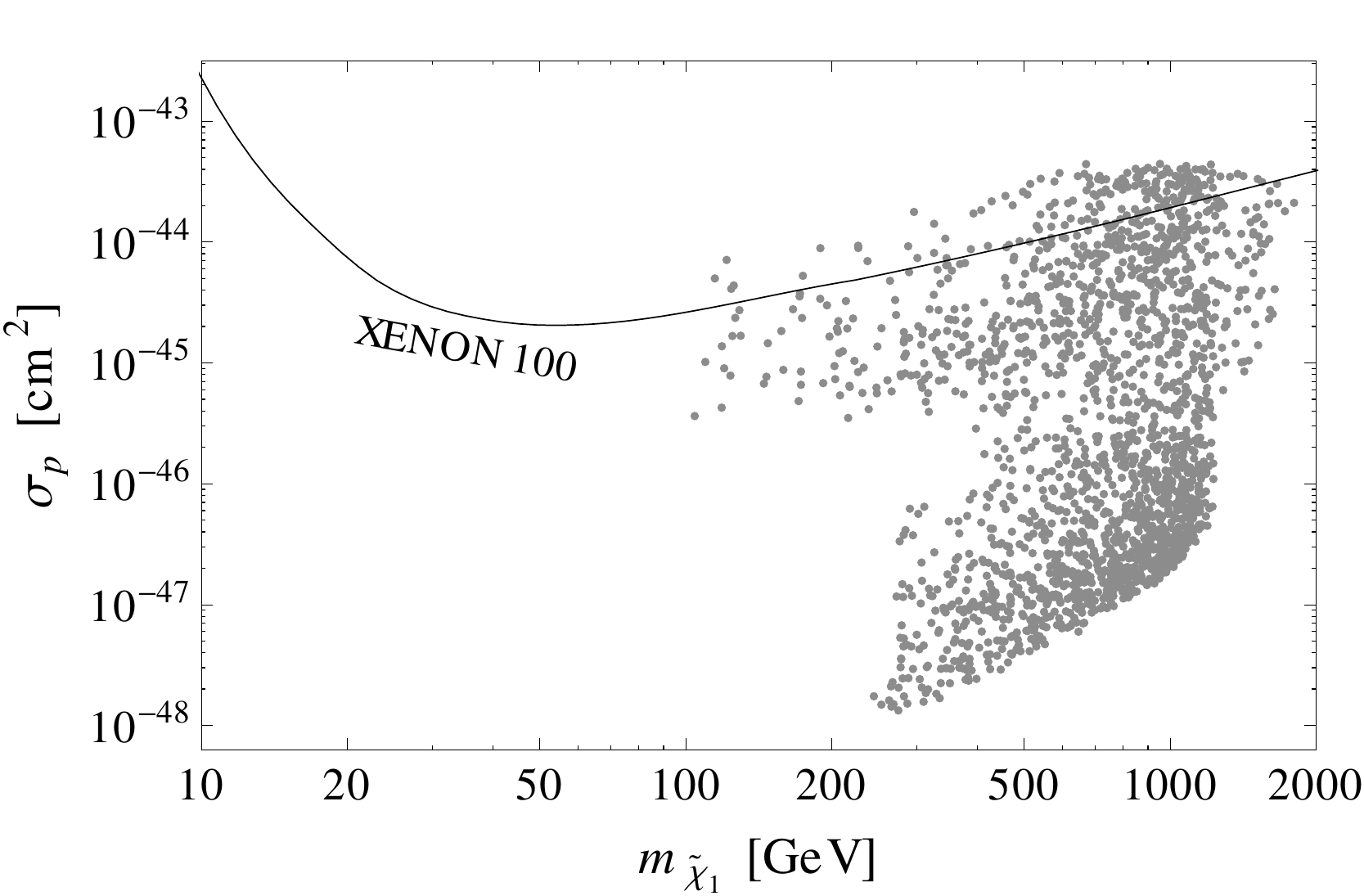}
\end{center}
\caption{\footnotesize{Neutralino proton cross section for the benchmark points with successful gauge coupling unification (gray). The current limit from the XENON100 direct dark matter search is also shown. The latter is only applicable if the lightest neutralino accounts for all dark matter in the universe.}}
\label{fig:directdetection}
\end{figure}

Turning to direct dark matter detection, we find that the cross section of the
lightest neutralino with nucleons is dominated by exchange of the light Higgs
(as the other scalars are in the multi--TeV regime). The coupling of the
lightest neutralino to the Higgs scales with the gaugino / higgsino mixing
angle, it vanishes in the limit of a pure state. The neutralino proton cross
section $\sigma_p$ for the benchmark points with PGU is shown
in \Figref{fig:directdetection}. For the determination of $\sigma_p$ we used
MicrOMEGAs~2.4.5, but took the updated nuclear quark form factors
from~\cite{Belanger:2013oya} suggested by recent lattice QCD calculations (see
also~\cite{Thomas:2012tg}). Note that these form factors suggest a smaller
strange quark contribution to the nucleon mass compared to earlier
computations~\cite{Pavan:2001wz}. They thus tend to give a smaller $\sigma_p$.

We find that $\sigma_p$ scatters between $10^{-48}\:\text{cm}^2$  and
$10^{-43}\:\text{cm}^2$. While the XENON100 experiment~\cite{Aprile:2012nq} has
just begun to probe this regime of cross sections, the next generation of direct
detection experiments will be able to test a significant fraction of the
depicted benchmark points. Still, there exist benchmark points with a bino--like
LSP which have a strongly suppressed $\sigma_p$. These may even be consistent
with thermal dark matter production due to coannihilations. Consequently, there
is no guarantee for a signal in direct dark matter detection. One should also
have in mind that the lightest neutralino must not necessarily account for the
dark matter in the universe such that direct detection experiments would be
doomed to fail.

\section{Conclusions}
\label{sec:conclusions}

This analysis is based on the hypothesis that precision gauge
unification (PGU) is not an accident. In general, imposing PGU reduces the
dimension of the soft supersymmetry breaking parameters by one. Within the
more `traditional' patterns of soft masses, imposing PGU implies typically a
large $\mu$ parameter, which is in conflict with usual `naturalness'
arguments. On the other hand,
a particularly attractive pattern of soft masses, consistent with PGU, is the one of
mirage mediation where the gaugino spectrum is compressed and the $\mu$ parameter
can be small. Such a spectrum can be realized in various string constructions which
exhibit a `competition' between gravity- and anomaly--mediated contributions to
soft masses. The similarity of gaugino masses at low energies  has profound
implications for collider and dark matter searches.

By requiring PGU we are driven in a corner of the MSSM
parameter space with a non--standard collider phenomenology. The small mass
difference between gluino and neutralino typically leads to events with a
reduced amount of visible energy. The detection of gluinos becomes far more
challenging than in most of the standard SUSY models. In the case of extreme
compression, initial state radiation is required to trigger on the SUSY events
and the detection efficiency is strongly reduced. The gluino is typically rather
long--lived, its decay length may exceed the LHC resolution and lead to
displaced vertices at the level of $\mu \text{m to mm}$. Even if
displaced vertices are absent, interesting information on the superpartner
spectrum can be obtained by studying the gluino decay modes.

Another virtue of compressed gaugino masses is that they generically lead to coannihilation
effects in the early universe. Cosmological problems related to the overproduction of dark matter
which are present e.g.\ in the CMSSM can easily be solved. Indeed, we find that in the parameter
space with PGU, the relic neutralino density is expected to be close to the observed dark
matter density.

Finally, it is needless to say that measuring superpartner spectra that
support PGU will provide us with invaluable information on how the standard
model is completed in the ultraviolet.

\section*{Acknowledgments}
We would like to thank Jamie Tattersall and Jan Hajer for useful discussions.
This work was supported by the SFB--Transregio TR33 ``The Dark Universe''
(Deutsche Forschungsgemeinschaft), the European Union 7th network program
``Unification in the LHC era'' (PITN--GA--2009--237920), the DFG cluster
of excellence Origin and Structure of the Universe by Deutsche
Forschungsgemeinschaft (DFG) and the ERC Advanced Grant project
``FLAVOUR''~(267104).
\newpage
\appendix

\section{Dependence of PGU on $\boldsymbol{\tan\beta}$ and $\boldsymbol{m_{3/2}}$}
\label{sec:moreplots}
\begin{figure}[htp]
\begin{center}
\includegraphics[width=7.4cm]{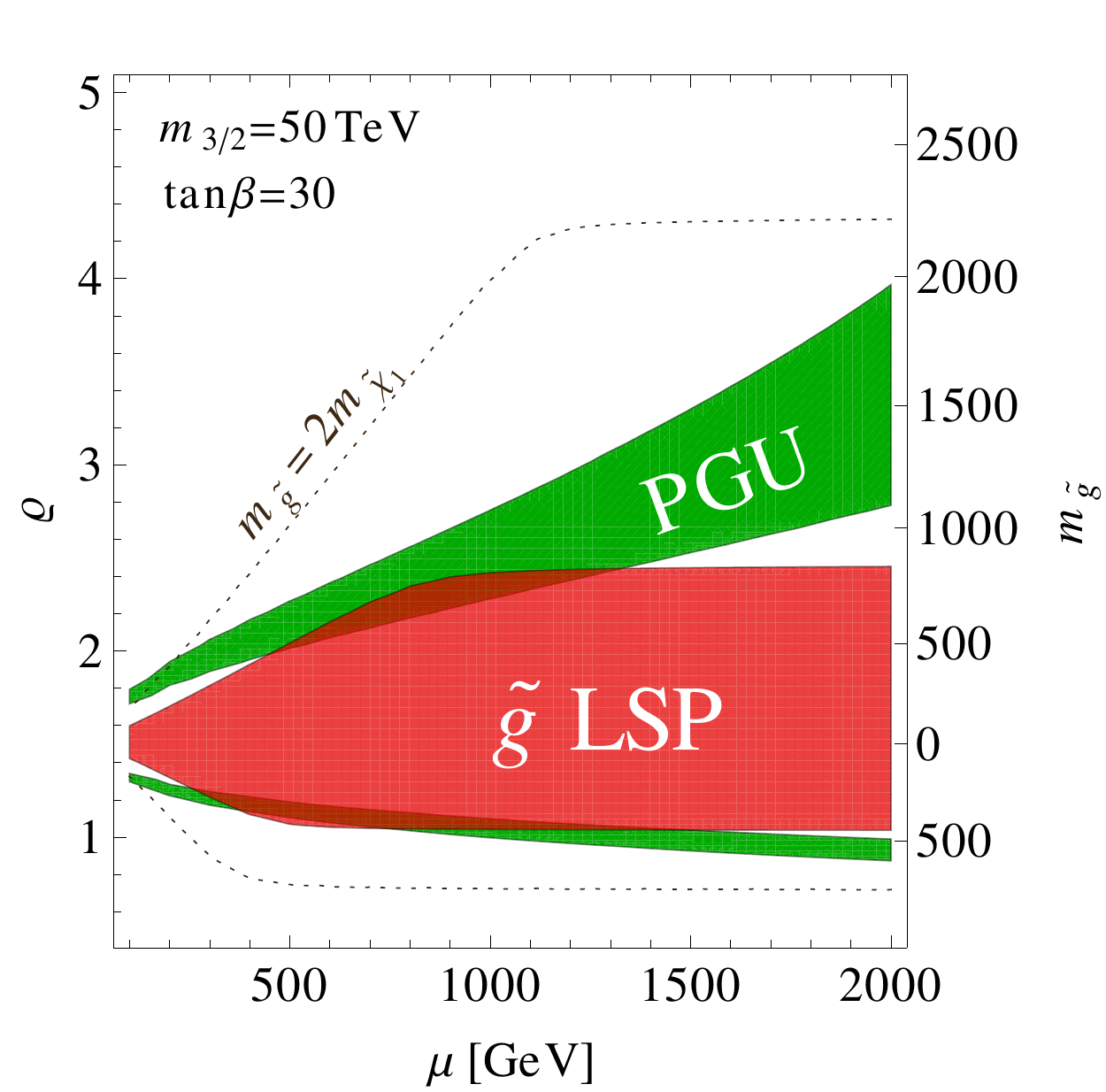}\hspace{5mm}
\includegraphics[width=7.4cm]{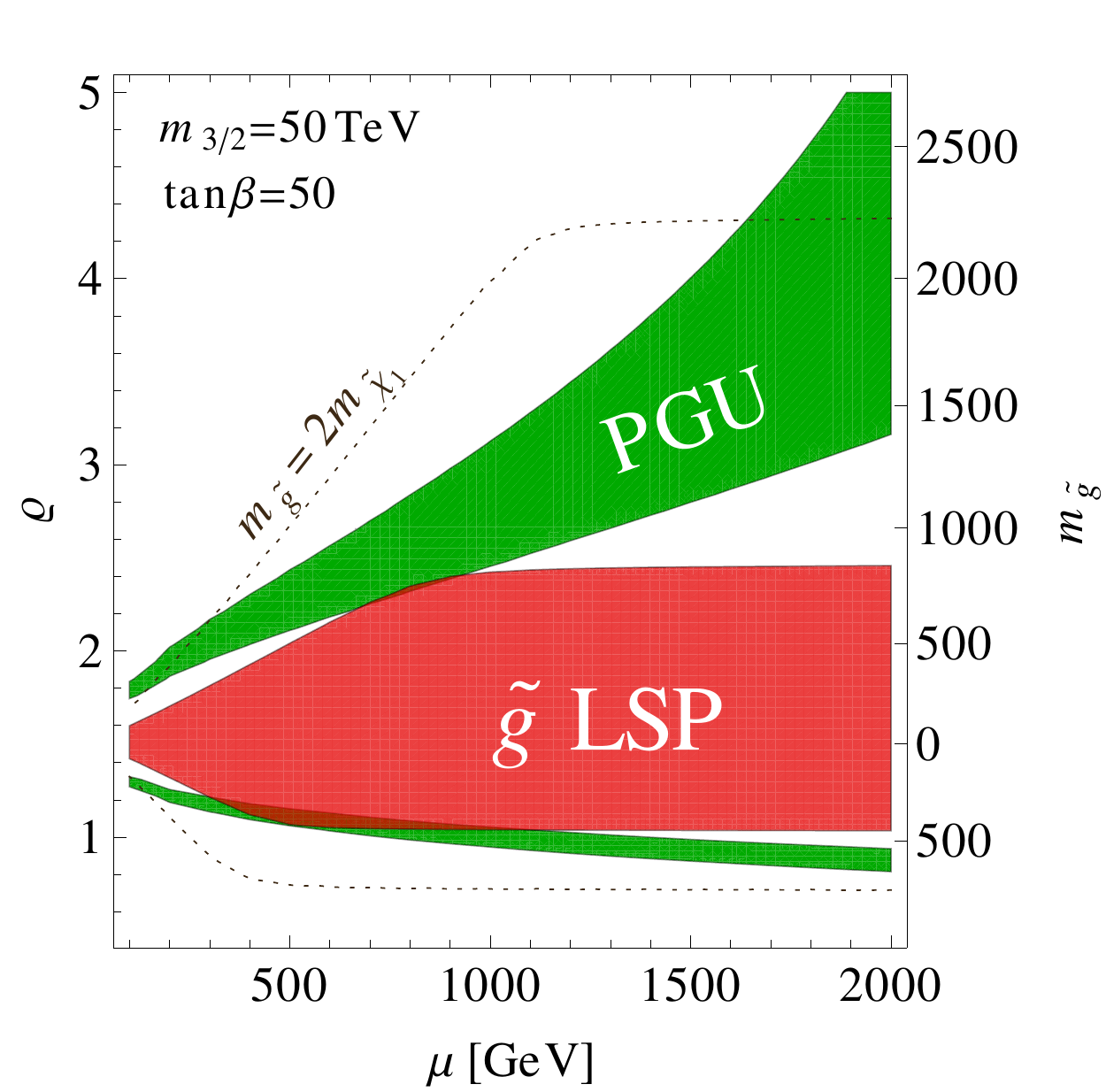}
\end{center}
\caption{\footnotesize{Same as figure~\ref{fig:Mirage} for different choices of $\tan\beta$.}}
\label{fig:varytanb}
\end{figure}

\begin{figure}[htp]
\begin{center}
\includegraphics[width=7.4cm]{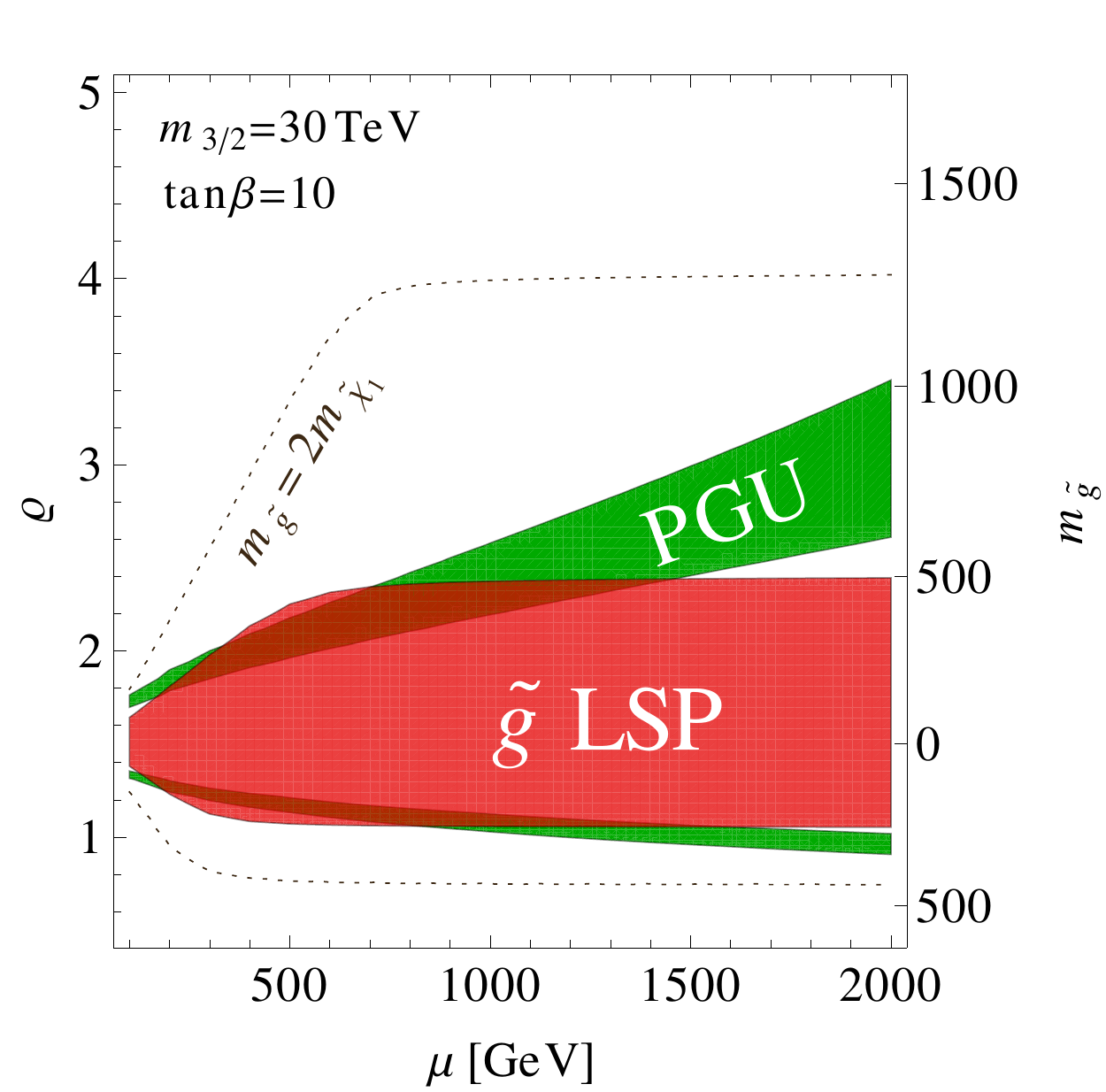}
\includegraphics[width=7.4cm]{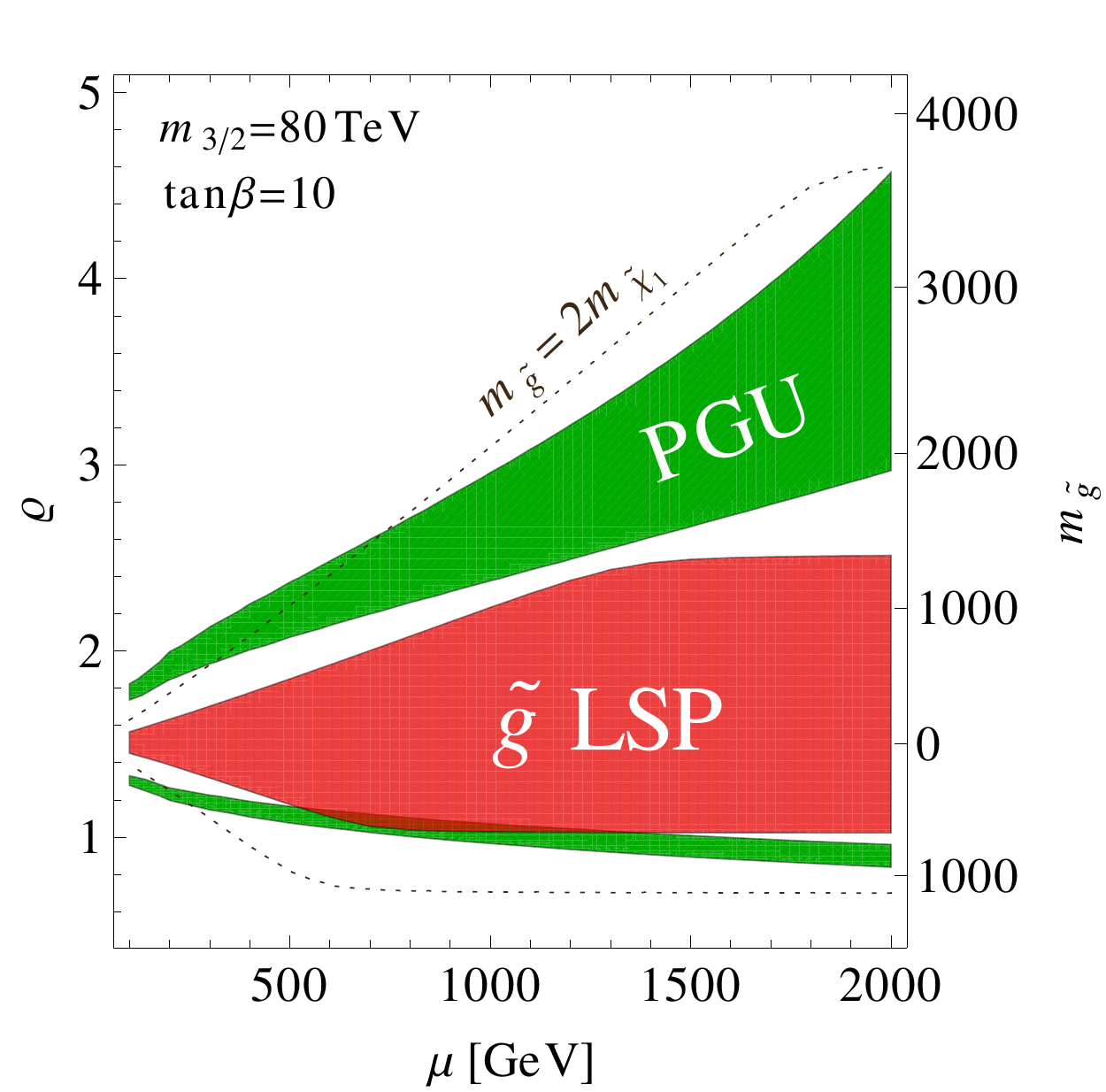}
\end{center}
\caption{\footnotesize{Same as figure~\ref{fig:Mirage} for different choices of $m_{3/2}$.}}
\label{fig:varym32}
\end{figure}

Figures~\ref{fig:varytanb} and~\ref{fig:varym32}  illustrate the dependence of
PGU on $\tan\beta$ and $m_{3/2}$.
\setlength{\abovecaptionskip}{-0.2cm}

\newpage
\bibliography{unification.bib}
\bibliographystyle{utphys2}
\end{document}